\documentclass[10pt]{article}
\usepackage{graphicx}

\def\Title#1{\begin{center} {\Large #1 } \end{center}}
\def\Author#1{\begin{center}{ \sc #1} \end{center}}
\def\Address#1{\begin{center}{ \it #1} \end{center}}

\newcommand\pubblock{\rightline{\begin{tabular}{l} Proceedings of the Fifth Annual LHCP\\ \pubnumber\\
         \pubdate  \end{tabular}}}

\newenvironment{Abstract}{\begin{quotation} \begin{center} 
             \large ABSTRACT \end{center}\bigskip 
      \begin{center}\begin{large}}{\end{large}\end{center} \end{quotation}}

\newenvironment{Presented}{\begin{quotation} \begin{center} 
             PRESENTED AT\end{center}\bigskip 
      \begin{center}\begin{large}}{\end{large}\end{center} \end{quotation}}

\def\beq{\begin{equation}}
\def\eeq#1{\label{#1}\end{equation}}
\def\eeqn{\end{equation}}

\def\beqa{\begin{eqnarray}}
\def\eeqa#1{\label{#1}\end{eqnarray}}
\def\eeqan{\end{eqnarray}}

\let\bar=\overbar

\def\Dslash{\not{\hbox{\kern-4pt $D$}}}
\def\dslash{\not{\hbox{\kern-2pt $\del$}}}

\def\msb{{\bar{\ssstyle M \kern -1pt S}}}

\usepackage{color}
\usepackage{hyperref}
\usepackage{lineno}

\newcommand\pubnumber{LHCb-PROC-2017-029 }

\newcommand\pubdate{\today}

\def\affiliation{
On behalf of the LHCb collaboration, \\
Bologna University and INFN, Bologna, 40126, Italy}

\begin{document}

\large
\begin{titlepage}
\pubblock

\vfill
\Title{  Novel strategies at LHCb for particle identification }
\vfill

\Author{ Fabio Ferrari  }
\Address{\affiliation}
\vfill
\begin{Abstract}

The LHCb experiment at the Large Hadron Collider (LHC) is performing high precision measurements in the flavour sector. 
An excellent performance of the particle identification (PID) detectors as well as the development of new data taking techniques are of fundamental importance in order to cope with increasingly harder challenges posed by the LHC Run 2.
The approach of data-driven calibration of particle identification performance at LHCb has changed significantly from Run 1 to Run 2 and calibration samples are now selected directly in the LHCb high-level trigger. 
This change of data-taking paradigm enables larger calibration samples with respect to Run 1 to be collected, giving access to low-level detector informations useful for studies of systematic effects, while retaining the same (or improving) the PID performances observed Run 1. 

\end{Abstract}
\vfill

\begin{Presented}
The Fifth Annual Conference\\
 on Large Hadron Collider Physics \\
Shanghai Jiao Tong University, Shanghai, China\\ 
May 15-20, 2017
\end{Presented}
\vfill
\end{titlepage}
\def\thefootnote{\fnsymbol{footnote}}
\setcounter{footnote}{0}
%

\normalsize 


\section{Introduction}
The LHCb detector~\cite{Alves:2008zz} is a single-arm forward spectrometer covering the \mbox{pseudorapidity} range $2 < \eta < 5$, designed for the study of particles containing $b$ or $c$ quarks.
Different types of charged hadrons are distinguished using information from two ring-imaging Cherenkov detectors.
Photons, electrons and hadrons are identified by a calorimeter system consisting of scintillating-pad and preshower detectors, an electromagnetic calorimeter and a hadronic calorimeter. Muons are identified by a system composed of alternating layers of iron and multiwire proportional chambers.
The trigger~\cite{Aaij:2014jba} consists of a hardware stage, based on information from the calorimeter and muon systems, followed by a software stage, which applies a full event reconstruction.

Almost all analyses performed on the data collected with the LHCb experiment ~\cite{Alves:2008zz} rely on particle identification (PID) information in order to discriminate between pions, kaons, protons, electrons, and muons. 
During Run 1, the only PID information available in the LHCb hardware and software triggers was that coming from the muon chambers. Since the beginning of Run 2, in 2015, also the information coming from the two Ring Imaging Cherenkov (RICH) detectors and the calorimeter (CALO) systems are included in the second level of the software trigger. These additions, coupled with the novel real-time detector alignment and calibration also introduced in Run 2, enable offline-quality PID variables in the trigger allowing for larger calibration samples with higher signal purity that can be used in physics analyses.

These samples are fundamental for studies relying extensively on PID variables as selection criteria. 
The data-driven methods used to assess PID performance are based on calibration samples. This can be achieved exploiting some kinematic properties of such decays. Moreover, such calibration samples need to be low multiplicity and the involved decays must be characterised by large branching fraction.

Starting from Run 2, data selected in the high-level software trigger (HLT) are persisted in the so-called \url{TURCAL} stream. Sufficient information is persisted from the software trigger such that no offline reconstruction is required. Nonetheless, the raw event is retained and an offline reconstruction is also performed. In this way, both the HLT and offline PID variables are available for the analysts. This is done in order to evaluate the performance of PID requirements applied both in the HLT and offline

\section{Centralized productions}
Calibration samples in Run 2 are composed by hundreds of million of events. These samples need to be accessed frequently and by many users that often require also RAW level information for their analyses. In order to satisfy these needs, starting from Run 2 the calibration samples have been produced centrally and made available to everyone. This approach goes under the name of \emph{working group production} (WGP). The data are written in two different formats: ROOT \url{TTree} (\emph{nTuples}), easy and fast to access, and $\mu$DST format, that contains only a sub-set of data of interest from a file in the standard LHCb DST format, and can be used to create \emph{nTuples} in order to perform specific studies. 

\begin{figure}[htb]
\centering
\includegraphics[width=0.9\textwidth]{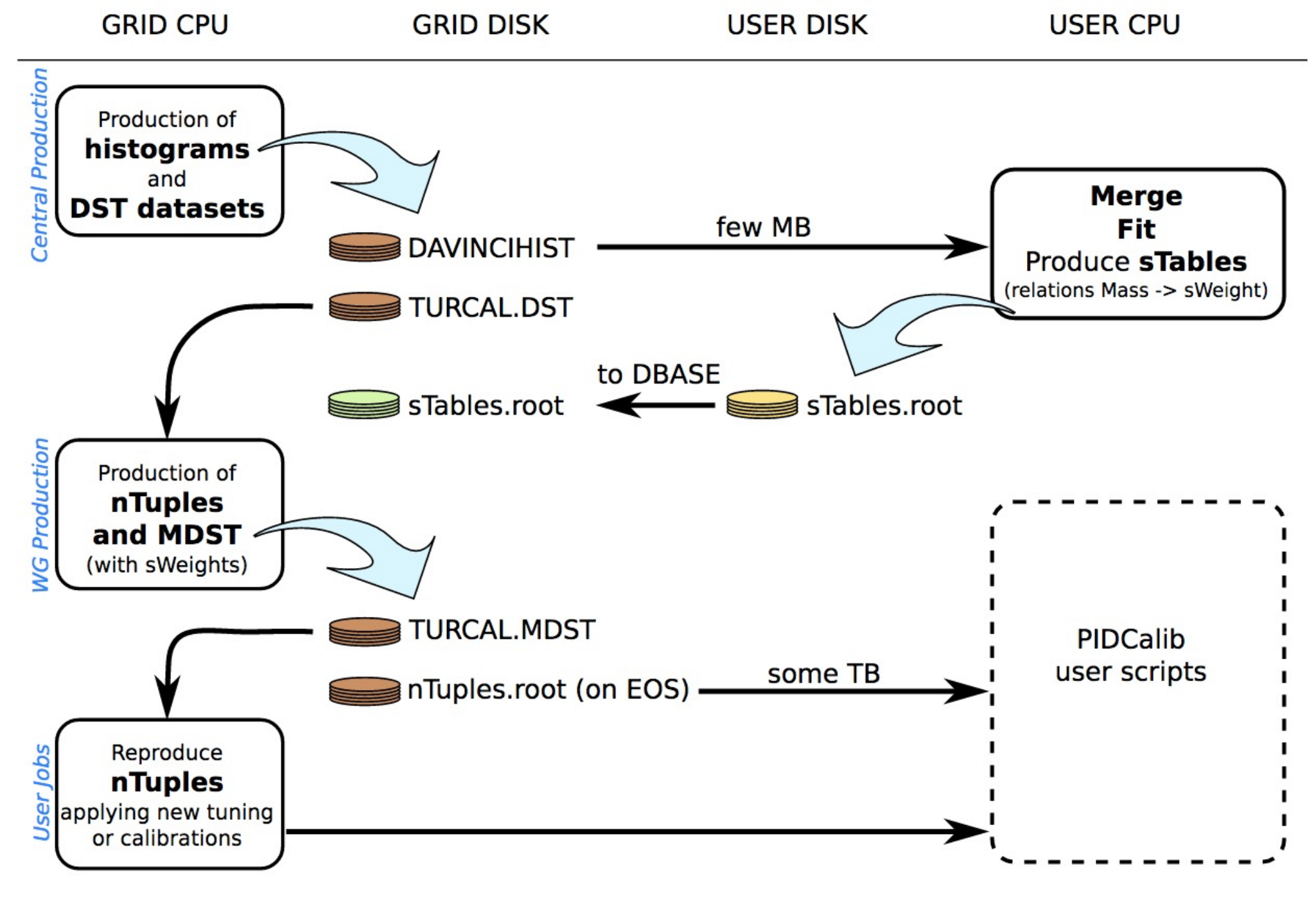}
\caption{Schematic representation of the data-flow of calibration samples.}
\label{fig:WGP}
\end{figure}

The data follow the flow sketched in Fig.~\ref{fig:WGP}. Firstly, data coming from the HLT are saved in histograms and DST datasets. 
Since the calibration samples need to be background-subtracted, the histograms are then fitted with the RooFit package~\cite{Verkerke:2003ir}, in order to obtain \emph{sWeights} that are then stored into a database accessible by user jobs on the GRID. 
Then, the full calibration sample, that is saved in the \url{TURCAL.DST} stream, is processed through a centralized production that pairs each candidate with the corresponding \emph{sWeight} and saves everything to both ROOT \url{TTree} and $\mu$DST formats. 
The produced \emph{nTuples} are then stored on a common repository (EOS), where they can be accessed by any user and by the official LHCb software, the PIDCalib package~\cite{PIDCalib}. 
If users want to reproduce the \emph{nTuples} in order to apply new tunings or calibrations, they are free to do so, since also the $\mu$DST format is stored on EOS and available to everyone.

\begin{figure}[htb]
\centering
\includegraphics[width=0.48\textwidth]{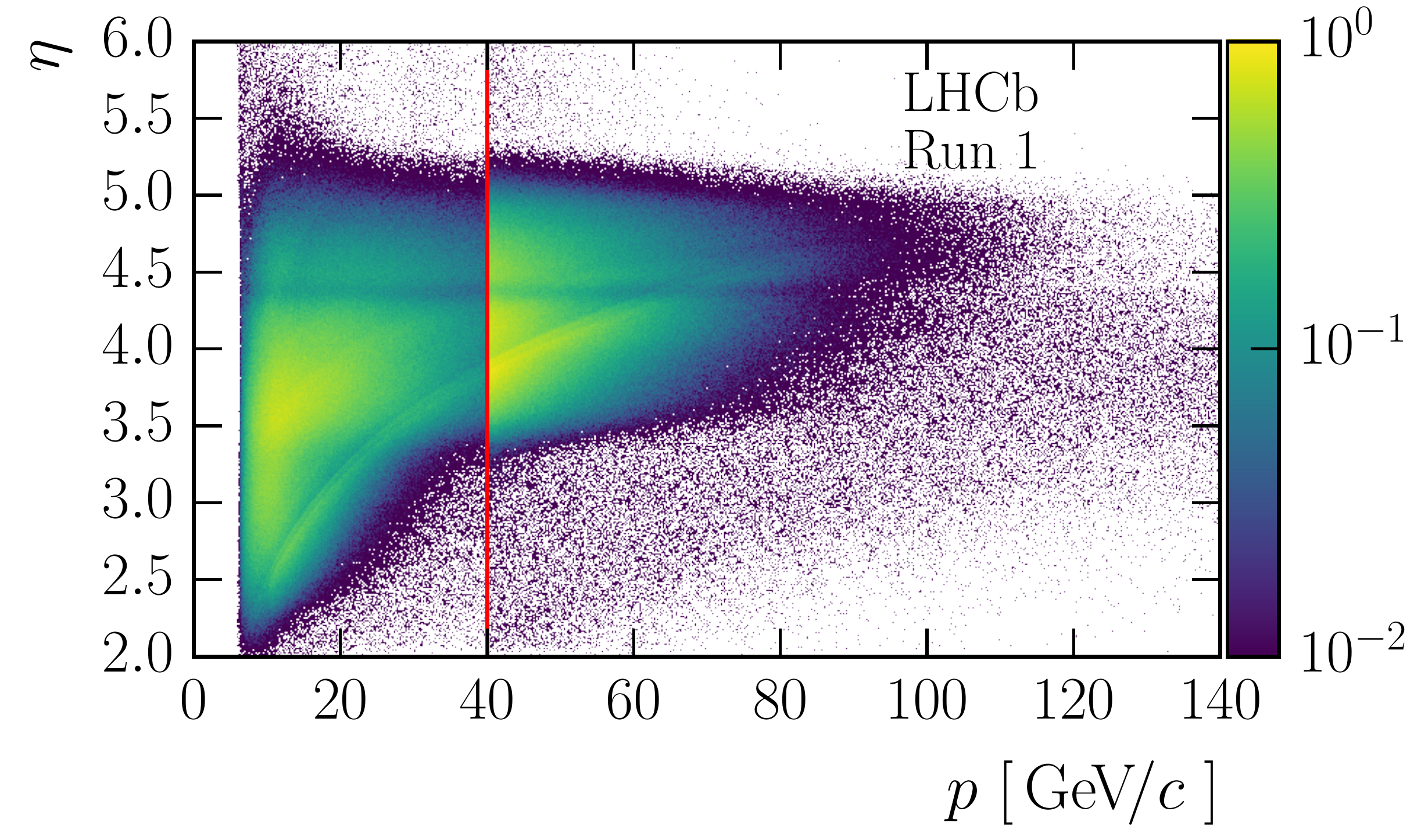}
\includegraphics[width=0.48\textwidth]{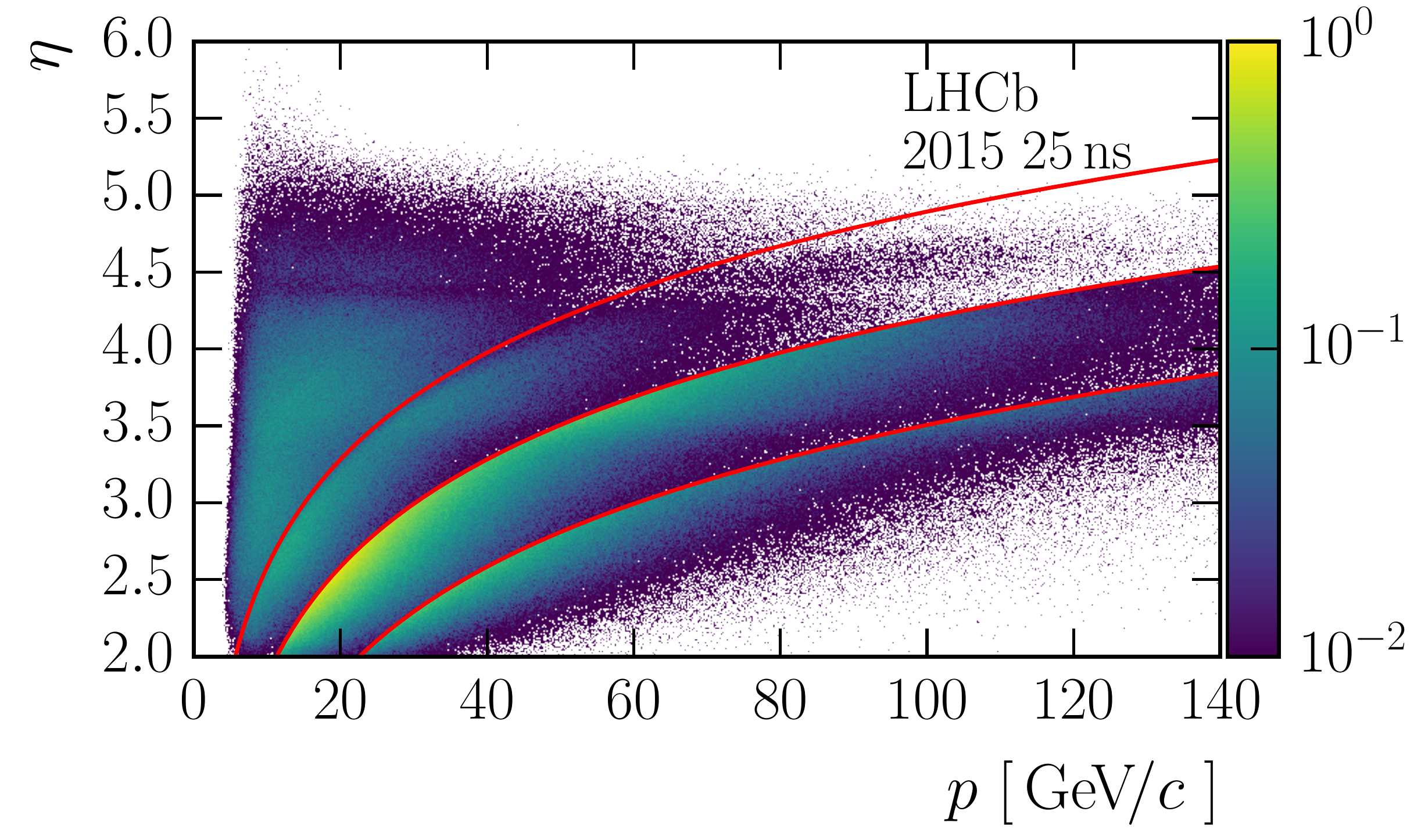}
\caption{Comparison of distributions between protons from (left) Run 1 and (right) Run 2 calibration samples in the $p-\eta$ plane. Note that the plots are normalised. The red lines separate samples selected by different trigger lines.}
\label{fig:petaPlaneComparison}
\end{figure}

Several new calibration samples have been added to those already selected during Run 1. In Tab.~\ref{tab:CalibrationSamples} the modes that are currently part of the calibration datasets are listed divided by species and $p_{\mathrm{T}}$ region. Owing to the new calibration samples, the phase-space coverage is significantly improved. As an example, we show in Fig.~\ref{fig:petaPlaneComparison} a comparison of the distributions in the $p-\eta$ plane for protons from Run 1 and Run 2.

\begin{table}[t]
\begin{center}
\begin{tabular}{c|c|c}  
Species & soft $p_{\mathrm{T}}$ & hard $p_{\mathrm{T}}$\\ \hline
$e^\pm$ & $-$ & $B^+ \to [e^+ e^-]_{J/\psi} K^+$ \\           
$\mu^\pm$ & $D^+_s \to [\mu^+ \mu^-]_{\phi} \pi^+$ & $J/\psi \to \mu^+ \mu^-$, $B^+ \to [\mu^+ \mu^-]_{J/\psi} K^+$ \\
$\pi^\pm$ & $K^0_S \to \pi^+ \pi^-$ & $D^{*+} \to [K^- \pi^+]_{D^0} \pi^+$ \\
$K^\pm$ & $D^+_s \to [K^+ K^-]_{\phi} \pi^+$ & $D^{*+} \to [K^- \pi^+]_{D^0} \pi^+$ \\
$p^\pm$ & $\Lambda^0 \to p \pi^-$ & $\Lambda^0 \to p \pi^-$, $\Lambda^+_c \to p^+ K^- \pi^+$, $\Lambda^0_b \to [p^+ K^- \pi^+]_{\Lambda^+_c} \mu^- \nu_{\mu}$
\end{tabular}
\caption{Summary of the modes collected in the calibration samples, divided by species and $p_{\mathrm{T}}$ regimes.}
\label{tab:CalibrationSamples}
\end{center}
\end{table}

\section{Run 2 PID performance}
Combining the informations from the PID sub-detectors, two sets of global PID variables are built.

\begin{itemize}
\item $\Delta \log \mathcal{L}(h-h^{\prime})$: this quantity is the change in the $\log \mathcal{L}$ between the $h$ and $h^{\prime}$ hypotheses for the particle under consideration.  
\item ProbNN: this quantity is the output of multivariate techniques created by combining tracking and PID information. This results in single probability values for each particle hypothesis.
\end{itemize}

\subsection{RICH}
LHCb RICH detectors provide excellent $p/K/\pi$ discrimination in the momentum range $2-100\ \mathrm{GeV}/c$ ~\cite{Adinolfi:2012qfa}. 
Since the two RICH are filled with different radiators $(C_4F_{10}$ and $CF_4)$, they provide discrimination in different momentum ranges.

\begin{figure}[htb]
\centering
\includegraphics[width=0.48\textwidth]{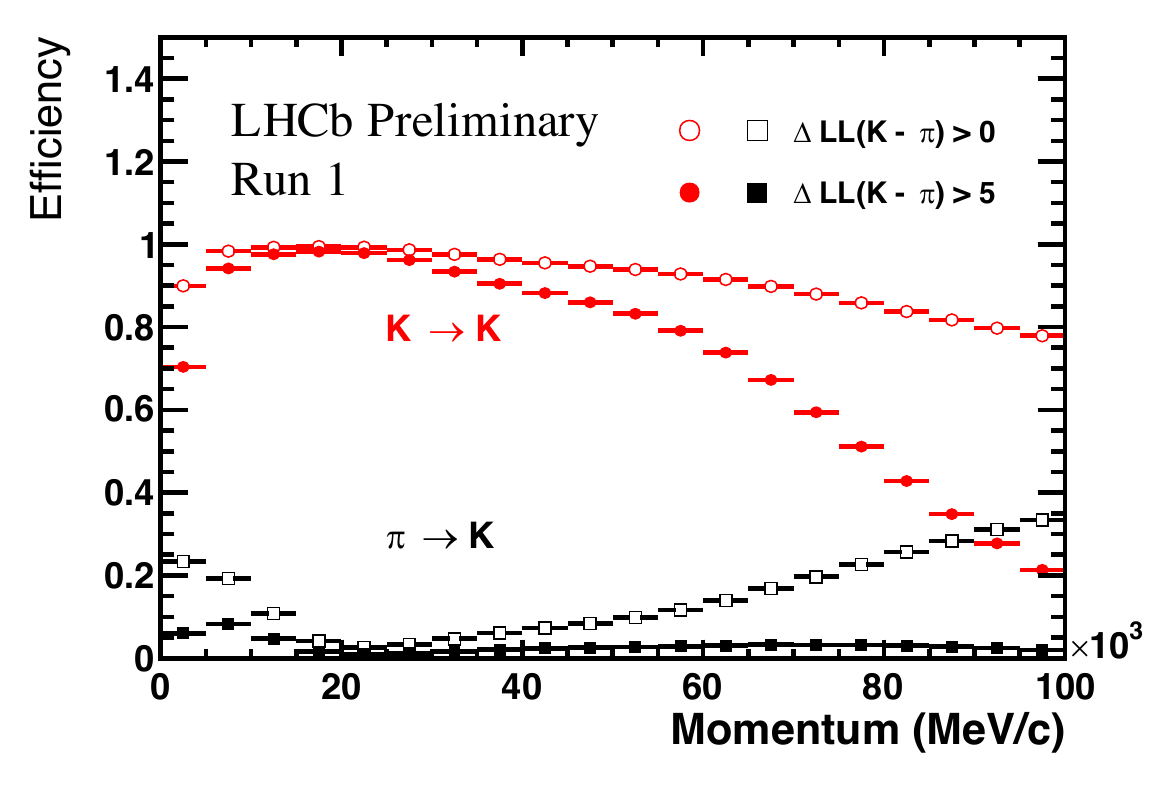}
\includegraphics[width=0.48\textwidth]{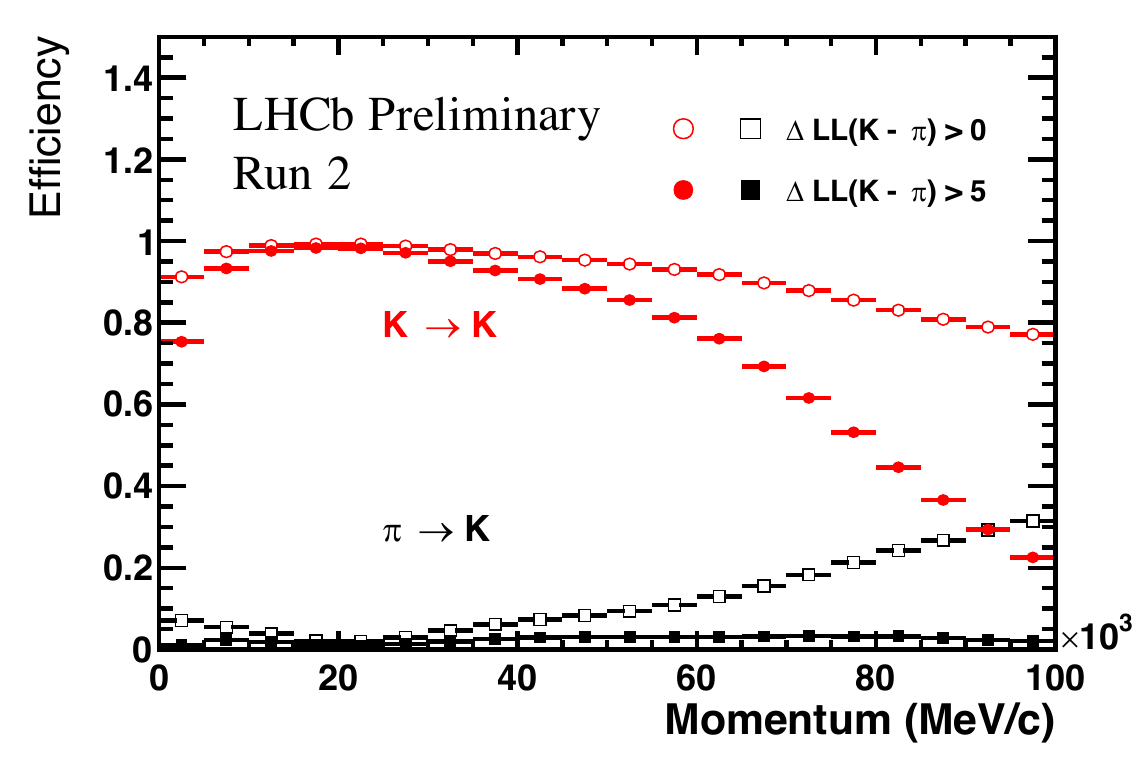}
\caption{Kaon identification efficiencies (red markers) and $\pi \to K$ mis-identification (black markers) probabilities during (left) Run 1 and (right) Run 2, for two different values of  $\Delta \log \mathcal{L}(K - \pi)$ requirement: (empty markers) looser requirement and (filled markers) tighter requirement.}
\label{fig:RICHComparison}
\end{figure}

We report in Fig.~\ref{fig:RICHComparison} the kaon identification efficiency and the $\pi \to K$ mis-identification probability, as a function of the particle momentum, separately for Run 1 and Run 2. 
As shown in the left plot, during Run 1 the LHCb experiment has achieved a $95\%(85\%)$ kaon identification rate with a $10\%(3\%)$ $\pi \to K$ mis-identification rate, applying as a reference the selection requirement $\Delta \log \mathcal{L}(K - \pi) > 0 (> 5)$. 
This performance has improved during Run 2 especially at low kaon momenta.While about the same kaon identification performance is kept, a decrease of the pion mis-identification probability is clearly visible.

\subsection{Muons}
\begin{figure}[htb]
\centering
\includegraphics[width=0.48\textwidth]{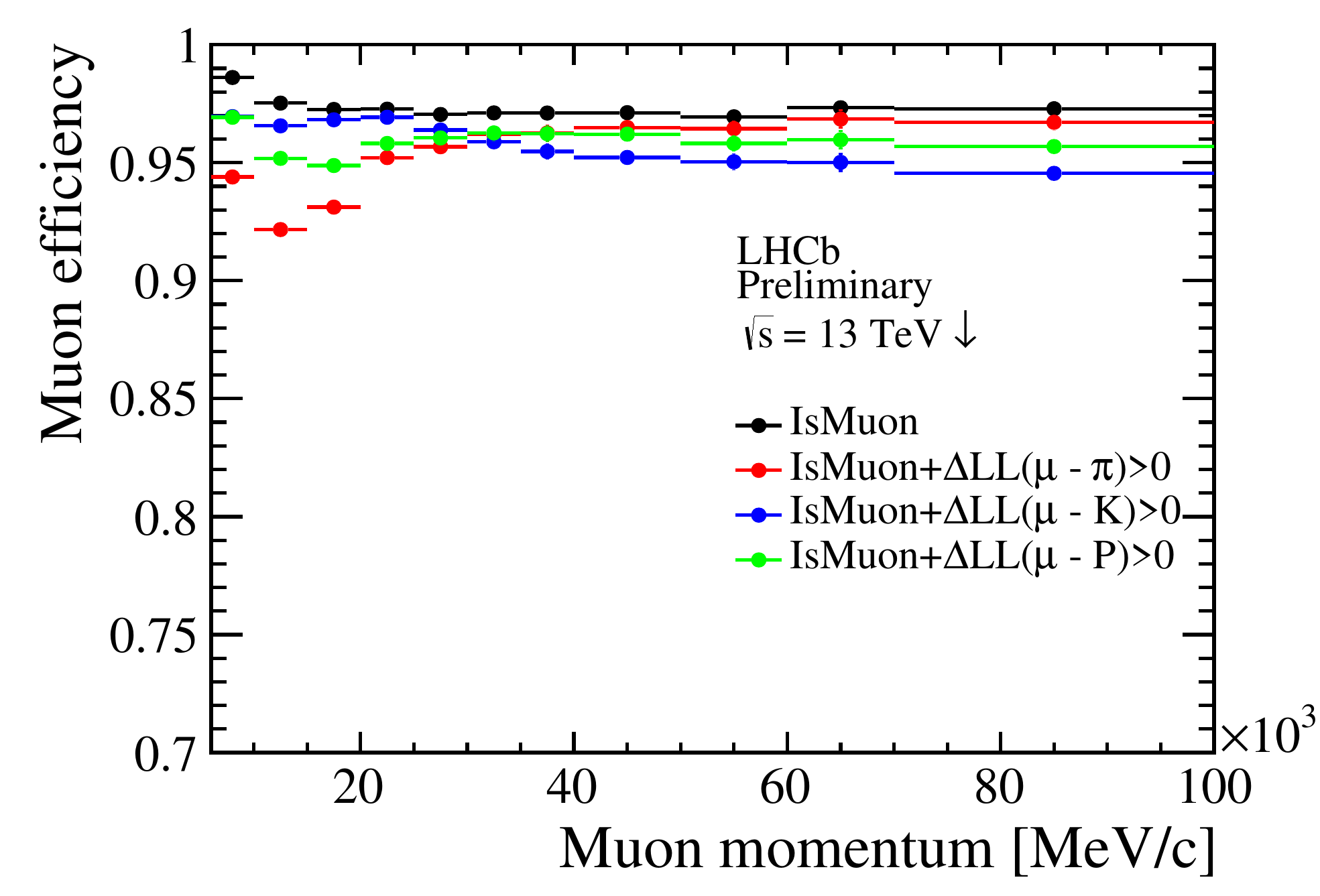}
\includegraphics[width=0.48\textwidth]{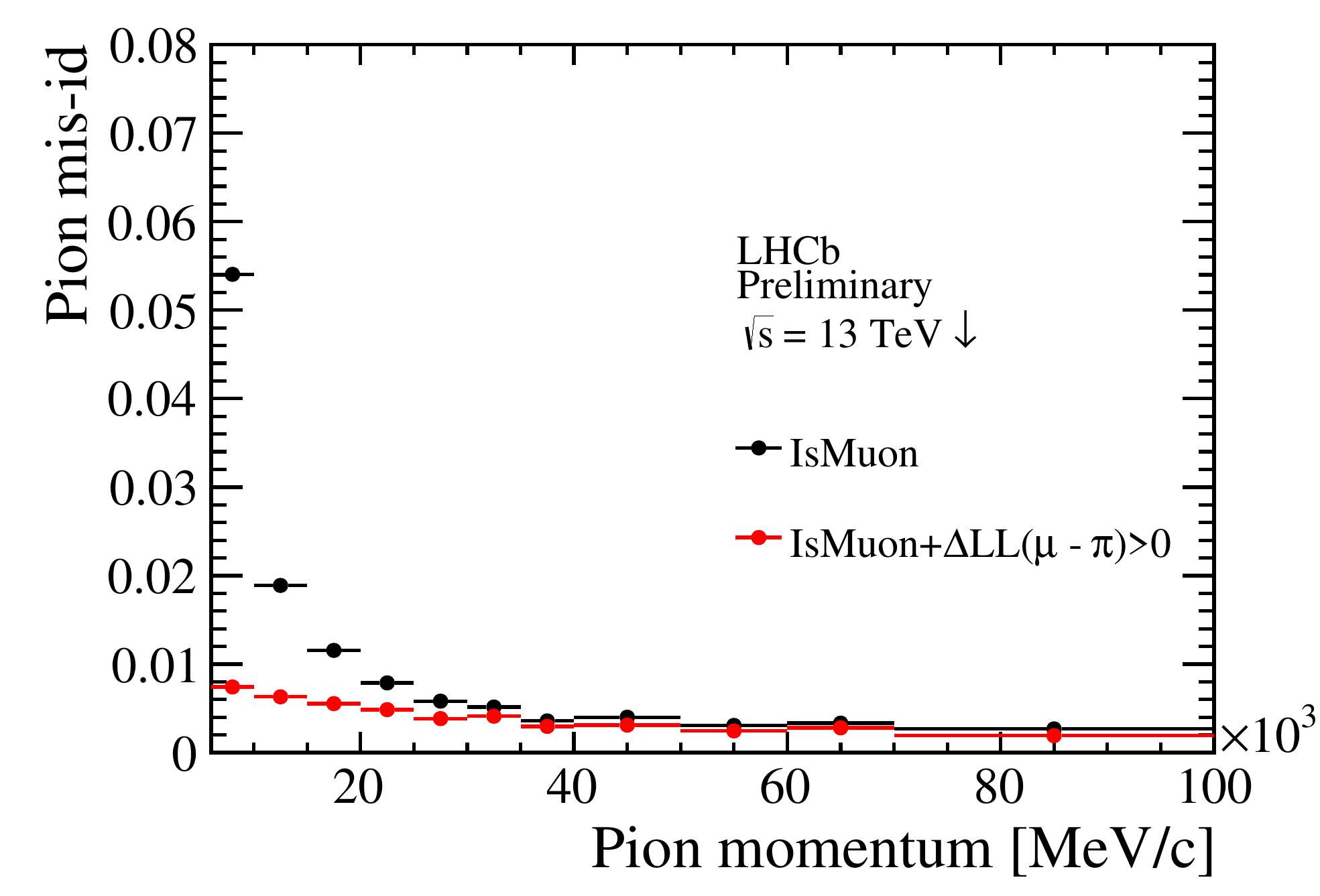}
\includegraphics[width=0.48\textwidth]{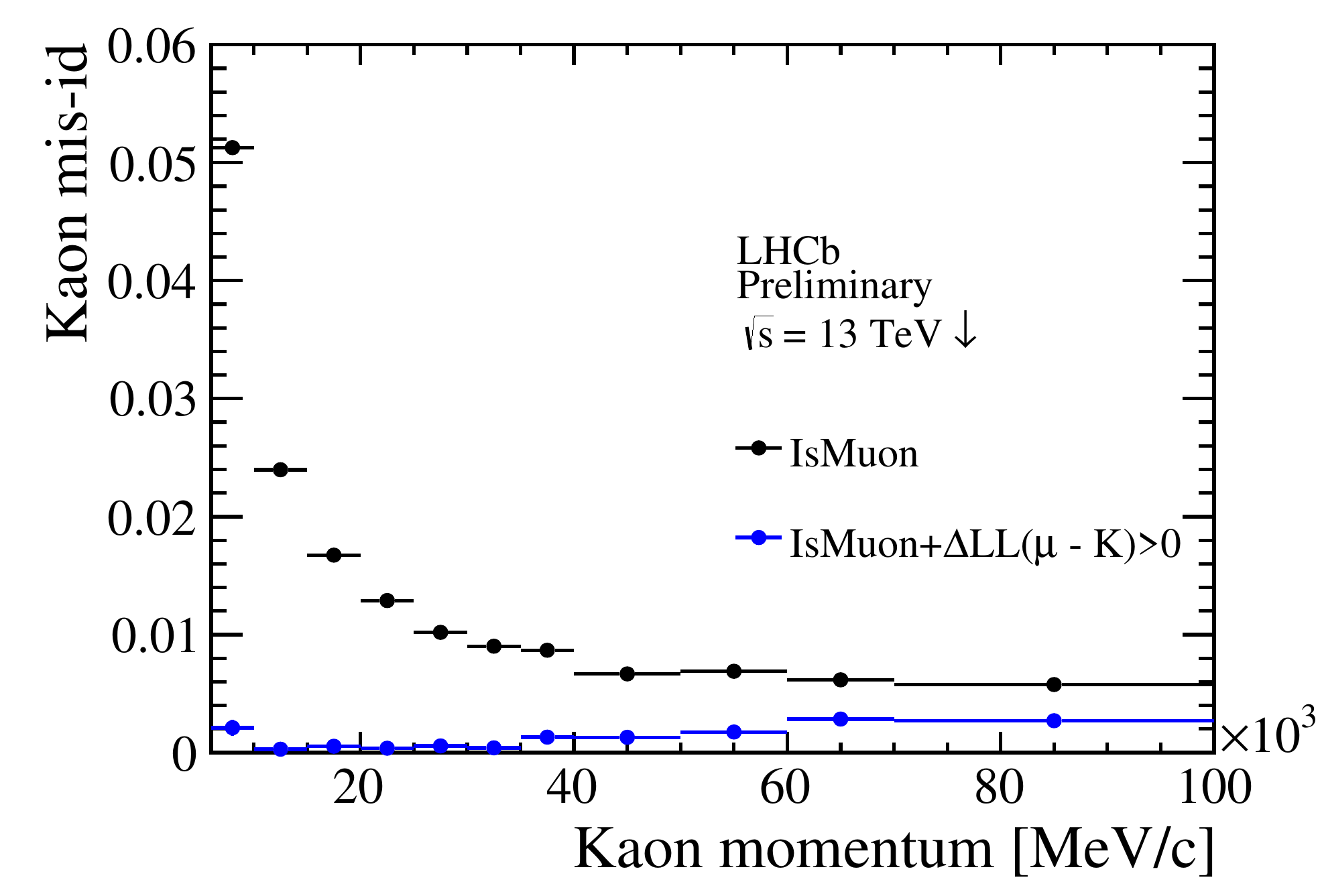}
\includegraphics[width=0.48\textwidth]{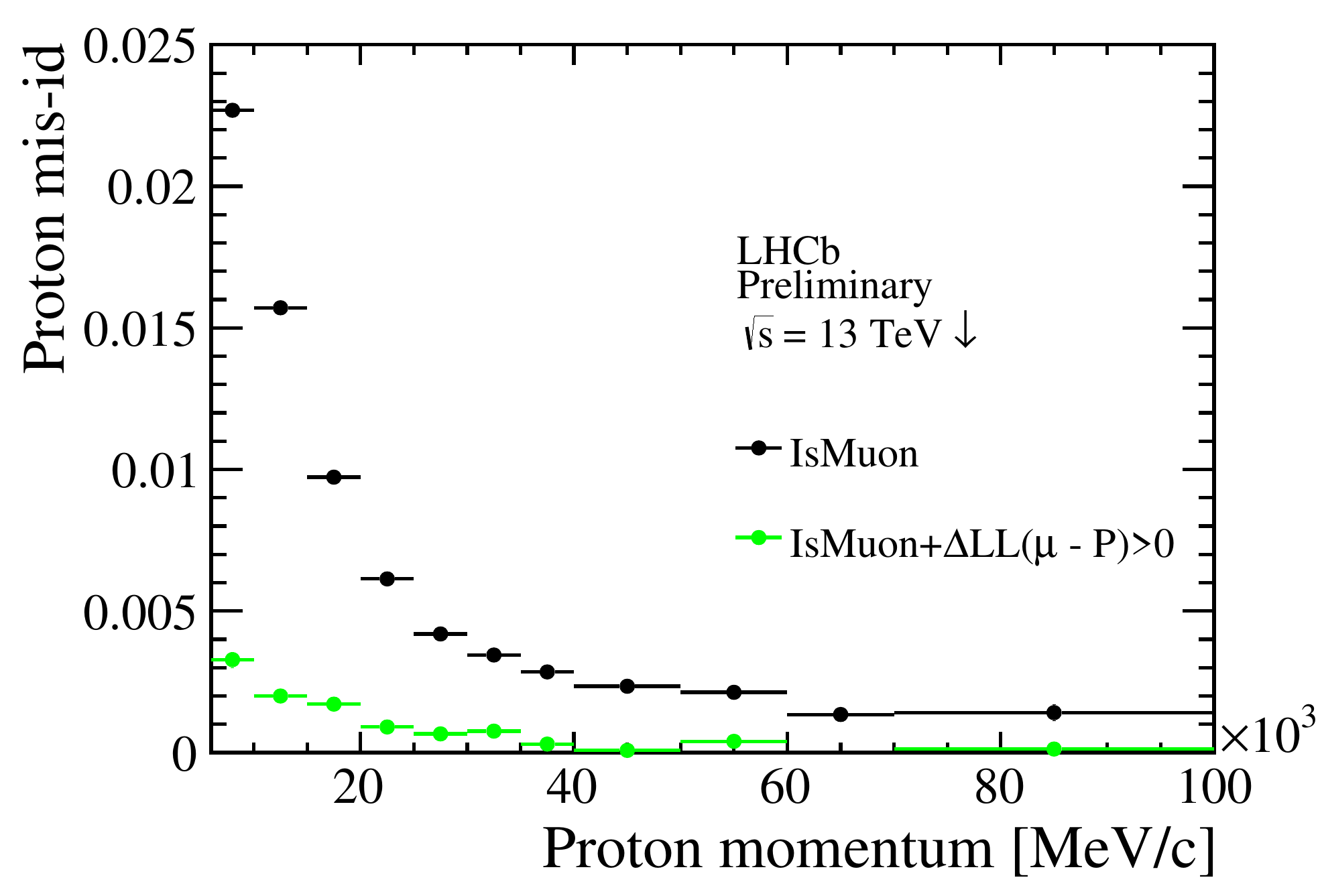}
\caption{Muon identification efficiencies (top left) and mis-identification probabilities (top right, bottom left, bottom right) for a common set of PID requirements. The calibration samples are from the 2015 datasets.}
\label{fig:MUONPerformance}
\end{figure}

The LHCb detector is equipped with five muon stations separated by layers of absorber, for a total of about $25$ hadronic interaction lengths~\cite{Archilli:2013npa}. 
The muon identification (IsMuon) is performed in different steps. 
Firstly, tracks are extrapolated to the muon stations. Then, an algorithm searches for hits around the extrapolated tracks and finally a probability is computed from the hit distribution in the muon stations. 
In Fig.~\ref{fig:MUONPerformance} the muon identification efficiency is shown for a set of different selection requirements aimed at rejecting hadrons and the mis-identification probability as a function of particle momentum for 2015 running conditions.

From the plots one can see that the integrated muon identification efficiency is about $95\%$, with a mis-identification probability generally lower than $1\%$ over the whole momentum spectrum. The identification and mis-identification rates are essentially the same as those obtained in Run 1 and reported in Refs.~\cite{Aaij:2014jba, Archilli:2013npa}.

\subsection{CALO}
The LHCb calorimeter system~\cite{LHCB:2000ab} discriminates between $e$, $\gamma$ and $\pi^0$.
We report in Fig.~\ref{fig:RICHComparison} the electron identification efficiency and the $\pi \to e$ mis-identification probability, as a function of the particle momentum for 2015 data.

\begin{figure}[htb]
\centering
\includegraphics[width=0.7\textwidth]{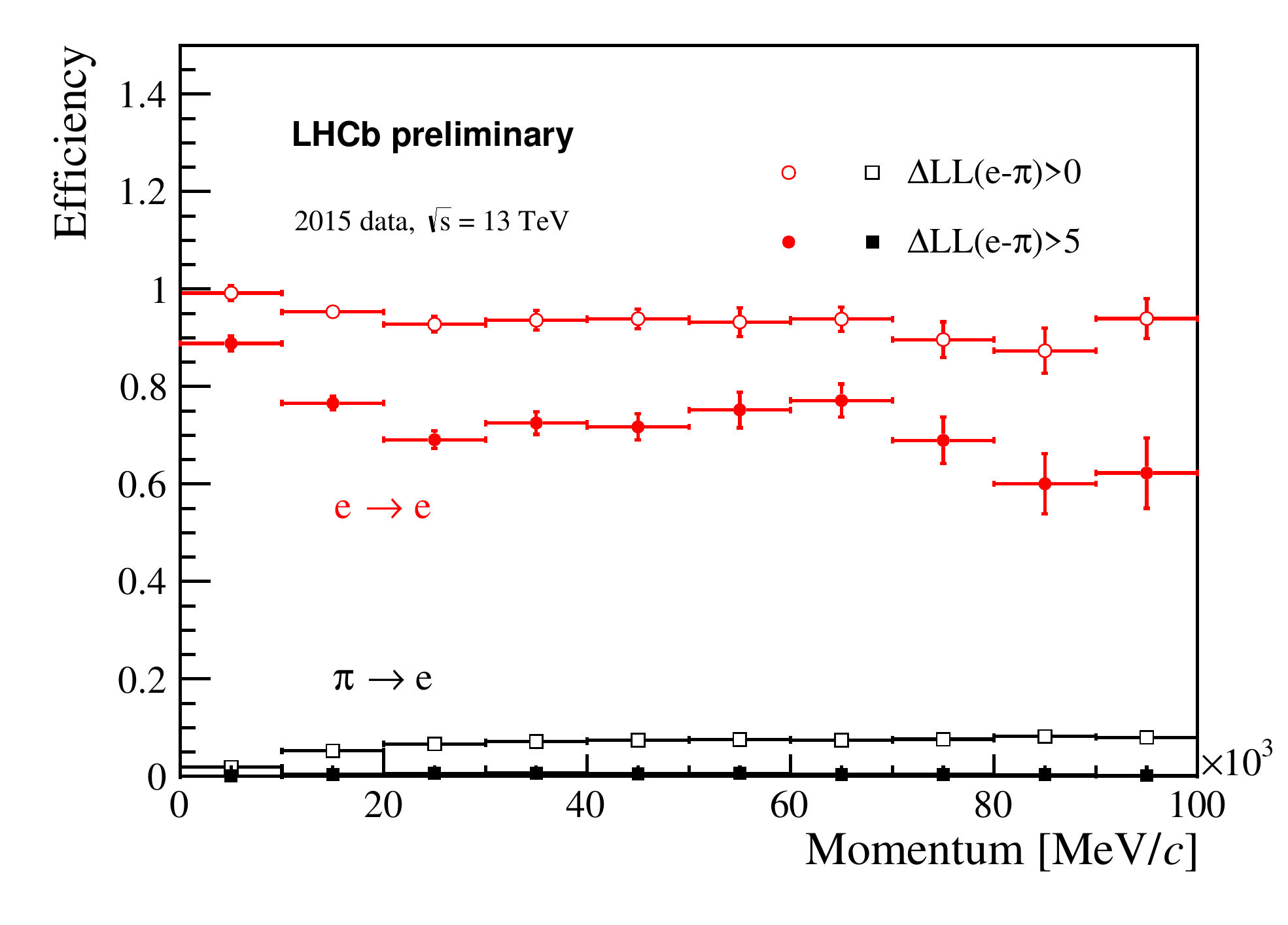}
\caption{Electron identification (red markers) and $\pi \to e$ mis-identification (black markers) for 2015 data, for two different requirements on $\Delta \log \mathcal{L}(e - \pi)$: (empty markers) looser requirement and (filled markers) tighter requirement.}
\label{fig:CALOComparison}
\end{figure}

By comparing the 2015 performance with that of Run 1 from Ref.~\cite{Aaij:2014jba}, it can be noted that the PID performance is essentially unchanged.

\section{Conclusions}
Starting from Run 2, calibration samples are selected directly in the HLT using offline-quality informations. This is possible due to a novel real-time detector alignment and calibration technique. 
It is now possible to select larger calibration samples with respect to Run 1, improving also the phase-space coverage by adding new decay modes. 
Moreover, it is also possible to employ low-level detector variables to perform systematic studies. All these features will be of fundamental importance also in view of studies for future LHCb detector upgrades.
As a result of all these novelties and enhancements, the LHCb Run 2 PID performance is at least at the same level as that obtained in Run 1, although the operating conditions are even more challenging.


\begin{thebibliography}{99}


\bibitem{Alves:2008zz}
  A.~A.~Alves, Jr. {\it et al.} [LHCb Collaboration],
  JINST {\bf 3} (2008) S08005.
  doi:10.1088/1748-0221/3/08/S08005

\bibitem{Aaij:2014jba}
  R.~Aaij {\it et al.} [LHCb Collaboration],
  Int.\ J.\ Mod.\ Phys.\ A {\bf 30} (2015) no.07,  1530022
  doi:10.1142/S0217751X15300227
  [arXiv:1412.6352 [hep-ex]].

\bibitem{Verkerke:2003ir}
  W.~Verkerke and D.~P.~Kirkby,
  eConf C {\bf 0303241} (2003) MOLT007
  physics/0306116.

\bibitem{PIDCalib}
  L.~Anderlini {\it et.al},
  LHCb-PUB-2016-021; CERN-LHCb-PUB-2016-021.

\bibitem{Adinolfi:2012qfa}
  M.~Adinolfi {\it et al.} [LHCb RICH Group],
  Eur.\ Phys.\ J.\ C {\bf 73} (2013) 2431
  doi:10.1140/epjc/s10052-013-2431-9
  [arXiv:1211.6759 [physics.ins-det]].

\bibitem{Archilli:2013npa}
  F.~Archilli {\it et al.},
  JINST {\bf 8} (2013) P10020
  doi:10.1088/1748-0221/8/10/P10020
  [arXiv:1306.0249 [physics.ins-det]].


\bibitem{LHCB:2000ab}
  [LHCb Collaboration],
  CERN-LHCC-2000-036.

\end{thebibliography}
\end{document}